# Pixelated Chemical Displays

Pierre-Alexandre Goyette[1], Étienne Boulais[2], Maude Tremblay[2], Thomas Gervais[1,2,3]

1  Institut de Génie Biomédical, École Polytechnique de Montréal, Montréal, QC, H3T 1J4, Canada
2  Department of Engineering Physics, École Polytechnique de Montréal, Montréal, QC, H3T 1J4, Canada;
3  Institut du Cancer de Montréal, Centre de Recherche du Centre Hospitalier de l'Université de Montréal, Montréal, QC, H2X 0C1, Canada

Correspondence and requests for materials should be addressed to T.G. (email: thomas.gervais@polymtl.ca).

**Abstract**

An increasing number of applications in biology, chemistry, and material sciences require fluid manipulation beyond what is possible with current automated pipette handlers, such as gradient generation, interface reactions, reagent streaming, and reconfigurability. In this article, we introduce the pixelated chemical display (PCD), a new scalable strategy to highly parallel, reconfigurable liquid handling on open surfaces. Microfluidic "pixels" are created when a fluid stream injected above a surface is confined by neighboring identical fluid streams, forming a repeatable flow unit that can be used to tesselate a surface. PCDs generating up to 144 pixels are fabricated and used to project "chemical moving pictures" made of several reagents over both immersed and dry surfaces, without any physical barrier or wall. Overall, this article sets the foundation for massively parallel surface processing using continuous flow streams and showcases new possibilities in both wet and dry surface patterning and roll-to-roll processes.

## Main text

### Introduction

Since the turn of the century, automated liquid handling (ALH) technology has become ubiquitous in academic and industrial laboratories and at the core of a revolution in the life sciences, chemistry, and material sciences[1,2]. Yet, from a technological point of view, in the vast majority of cases, ALH embodiments remain a mere robotization of the human pipetting movement. ALH systems can deposit small volumes of liquid extremely efficiently using multiple pipette heads simultaneously. However, they cannot expose surfaces to gradients of chemicals, nor can they replace medium without dilution or slow intermediate rinsing steps[3]. There is, meanwhile, an increasing demand for systems with a faster spatiotemporal control of chemical species, which can be explained by a growing interest in cellular pathway kinetics by both the synthetic biology and pharmaceutical domain[4,5].

In an attempt to free the life sciences from the "tyranny of pipetting" [6], significant developments in the field of open-space microfluidics have brought new strategies for ALH compatible with well-based or petri-based assays. These include, among others, the microfluidic pipette[7], the chemistrode[8], the microchemical pen[9], and perhaps more importantly, the microfluidic probe[10]. All are contact-free methods which operate "from above" by continuously streaming reagents above an immersed surface. They can switch between two or more reagents quickly without dilution, enhance convective mass transport to surfaces, and generate sharp floating concentration gradients[7]. These technologies, however, are still inherently serial and slow. For example, the scanning speed of reagent-streaming devices is limited by the incubation time of the reagents involved in the surface reaction, often minutes to hours for protein-protein reactions (immunoassays) or cell stimulation (e.g. drug incubation). Finally, all open-space microfluidic systems developed so far can only operate on immersed surfaces since, for fluid confinement to occur, more fluid must be aspirated from the probe than what is injected, with the excess taken from the immersion medium[11–13].

Circumventing limitations to both standard and microfluidic-based ALH requires a radical perspective change from inherently serial scanning methods to massively parallel, reconfigurable reagent dispensing. The concept of very large scale microfluidic reagent processors eloquently demonstrated by Thorsen *et al*. almost two decades ago offer a path forward[14]. In this paper, we introduce the Pixelated Chemical Display (PCD), the ALH analog of a liquid crystal display (LCD), where a large number (100+) of independent reagent streams can be colocalized above a surface and reconfigured over time with discrete spatiotemporal control. We developed an exact first principles model to characterize mass transport under PCDs and provide experimental and theoretical evidence of their large-scale stability and robustness under flow perturbations or device imperfections. Devices are fabricated via a simple monolithic 3D printing process and used to demonstrate reagent confinement over both immersed and dry surfaces. In doing so, we challenge for the first time the received idea that open-space microfluidic devices must be operated immersed in fluids for reagents to be confined. Overall, the technology suggests a novel way to dynamically process surfaces by projecting "chemical moving pictures" onto them, which we exploit to propose broad applications in robotic liquid handling, roll-to-roll printing, and surface engineering.

**Pixelized chemical display concept**

The key to develop a massively parallel open-space microfluidic system is to define a basic modular flow subunit, or microfluidic pixel, that can be indefinitely repeated over a surface completely. Microfluidic pixels are a special type of polygonal microfluidic multipoles[15], where an injection is centered between three or more aspiration apertures that form the vertices of a polygon. Fluid is continually streamed through these apertures within the shallow gap formed between the device and the surface to be processed (Fig. 1). Adjacent microfluidic pixels share aspiration apertures on their vertices to form a lattice of apertures which, for certain regular polygonal shapes, can completely tessellate a surface with independent and isolated fluid confinement areas (Fig. 1). The architecture of pixelated devices is highly modular as any number of columns or line of microfluidic pixels can be added on the periphery without affecting the inner units. Inside a single pixel, reagents are transported by convection but also diffuse, to a certain extent, in neighboring pixels in the lattice. The flow rate within a pixel and the diffusivity of the tracer (reagents, particles, heat) will determine the amount of interpixel transport taking place. Based on previous recent mathematical developments describing convection-diffusion transport problem in planar viscous flows (or Hele-Shaw flows[16] and their early translation into microfluidics[15,17], we found an exact analytical solution for both streamlines and concentration profiles within pixels (SI section 2). We used this model throughout this paper as an experimental design tool to either minimize or exploit diffusional interfaces at the edge of the pixels in surface processing applications. Match between experiments and theory is outstanding for both diffusion and convection (SI section 2, Supplementary Fig. 7).

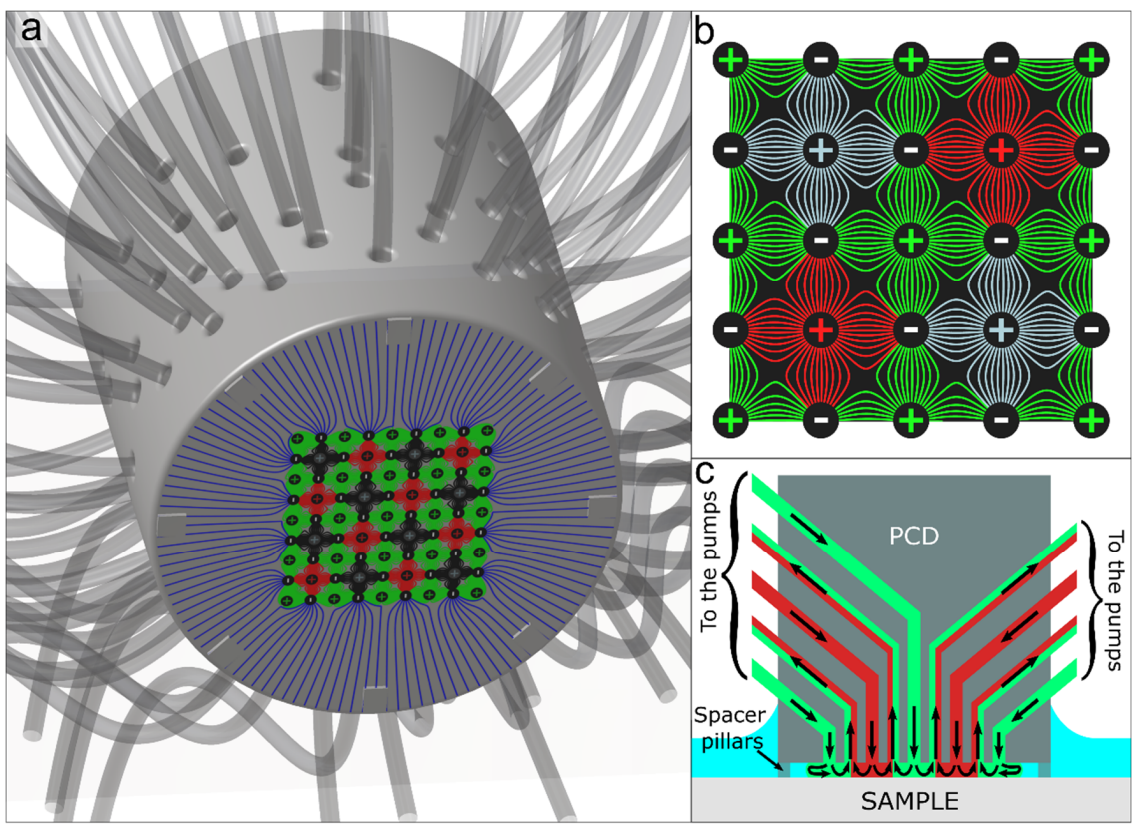

**Fig. 1.** Pixelated Chemical Display concepts and schematic. a) Schematic presenting an 81-aperture square-pixel PCD. The PCD is held over a sample. Lines on the surface represent the flow lines. b) Theoretical flow lines in an array of microfluidic pixels under a PCD. Positive and negative signs represent injection and aspiration apertures, respectively. c) Side view cut schematic of a PCD. The PCD is maintained at a fixed height over a surface with the help of spacer pillars. Fluids are injected and reaspirated within the shallow gap formed between the surface and the device.

## Device operation (457 words)

We tested PCDs based on the only three possible regular tessellation patterns in planar 2D space (square, triangle, and hexagon). An 81-aperture rectangular array and a 92-aperture staggered array PCDs were 3D-printed using stereolithography and a method previously developed[18]. To limit the number of pumps required, we designed custom-built manifolds to deliver fluids from one pump into all pixels sharing the same reagents. We also merged aspirations in two independently controlled groups (Supplementary Fig. 1), such that a PCD can be operated using a minimum number of pumps (2 + X, where X is the number of different reagents injected) no matter its size. Tubes connecting the manifolds to the PCDs were used as precision hydraulic resistors to match the flow rate from all apertures. Micrographs were taken using a fluorescence microscope through a glass slide that served as the sample surface (Supplementary Fig. 2). The three pixel shapes tested provided highly regular patterns (Fig. 2).

In 2D periodic lattices, where every pixel shares a no flux boundary condition with its neighbors, having a zero net flow rate is the only method to preserve local mass balance. An exception is possible for the outermost pixels, where edge effects are present. A PCD with the edge pixels having a negative net flow rate, such as in Fig. 2a-c, operates in the well-established hydrodynamic flow confinement regime[11,19]. For any aperture configuration other than microfluidic pixels, including every previously published open-space microfluidic system, this regime is the only possible way to confine streaming reagents since, without a net aspiration from the device, they would diffuse to infinity and escape confinement. A restrictive condition on their operation ensues, which is that devices in hydrodynamic flow confinement must necessarily be immersed in a fluid to operate. In contrast, the unique 2D periodic symmetry of PCDs preserves confinement of the inner pixels even in the case of a net outward flow rate (Fig. 2d-f). In other words, reagents streamed are locally confined in a globally unconfined flow. The main advantage of this operation mode is that PCDs no longer have to be used on an immersed surface. They can be used on dry surfaces, and also operated in an upside-down, fountain-like configuration. By injecting a neutral solution in the edge pixels, localized treatments of initially dry surfaces can thus be achieved (Supplementary Fig. 3). Another advantage of PCDs is that they are modular and can be made of any size. The pressure drop inside pixels is decoupled from the size of the system. Furthermore, the larger the system is, the more ideal the periodicity conditions on the inner pixels become. We fabricated PCDs comprising up to 144-pixel (313 apertures) and found that they provide very stable and regular flow patterns (Fig. 2g, h).

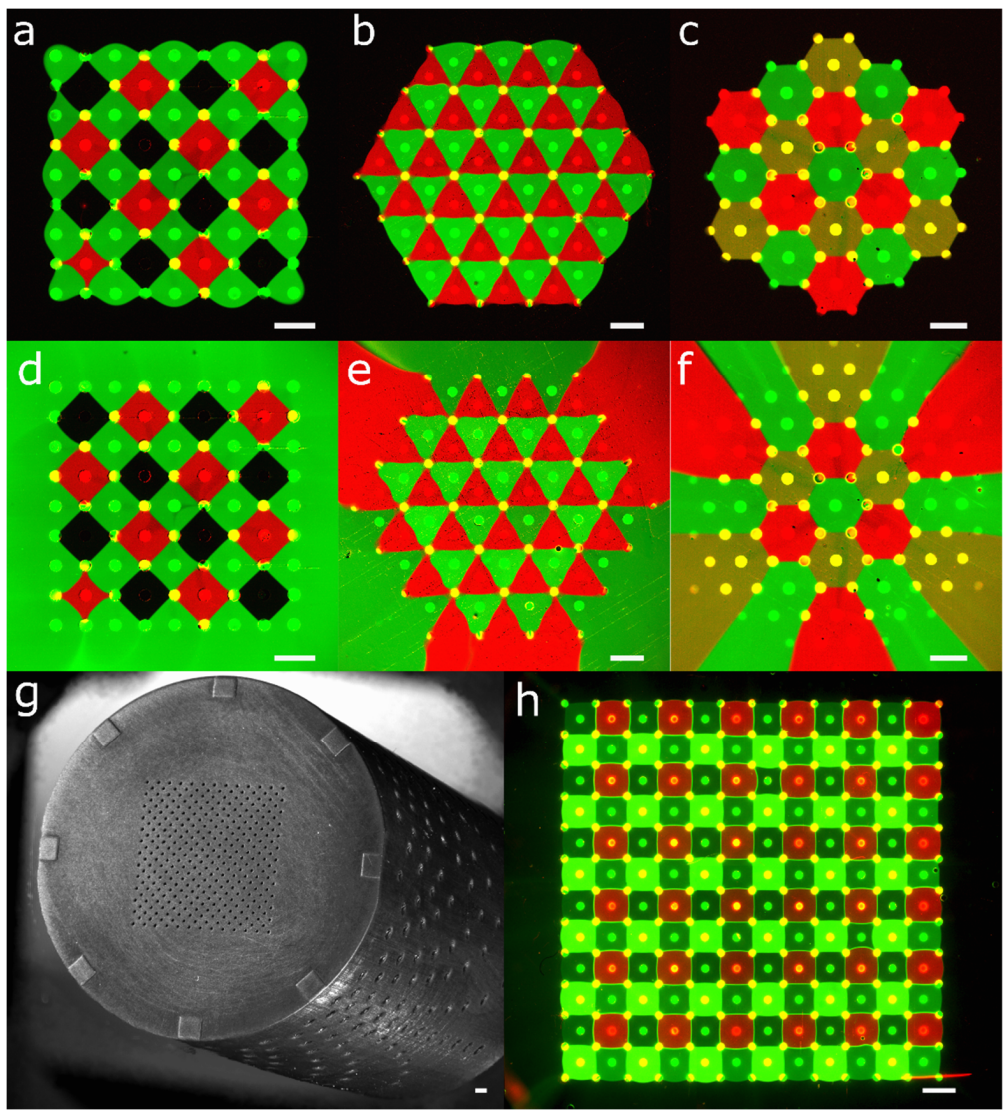

**Fig. 2.** Experimental PCDs. Fluorescent micrograph at the tip of a) a square-pixel PCD, b) a triangle-pixel PCD, and c) a hexagonal-pixel PCD. Fluorescent micrographs of a locally-confined, globally unconfined d) square-pixel PCD, e) triangle-pixel PCD, and f) hexagonal-pixel PCD. g) Micrograph of a 3D-printed 144-pixel PCD device. h) Fluorescent micrograph of a 144-pixel PCD. All the scale bars represent 1 mm.

**Stability**

PCD stability is crucial for device operation and was further investigated theoretically and experimentally. To ensure that PCD function is not compromised by a small number of clogged apertures, we compared transport inside a perfect PCD with that of a PCD with a single clogged injection. Both experimental results and numerical simulations showed that a clogged injection has only a small local effect on PCD operation (Fig. 3). Globally, microfluidic pixels are slightly shifted toward the errors with an amplitude decreasing quickly following a steep power law, such that even neighboring pixels are only mildly affected (Fig. 3a-c). These results are also in accordance with our experimental observations (Fig. 3d-e). This situation is again analogous in LCDs, where a "dead" pixel does not affect significantly its neighbors. We obtained similar results for clogged aspiration (Supplementary Fig. 4).

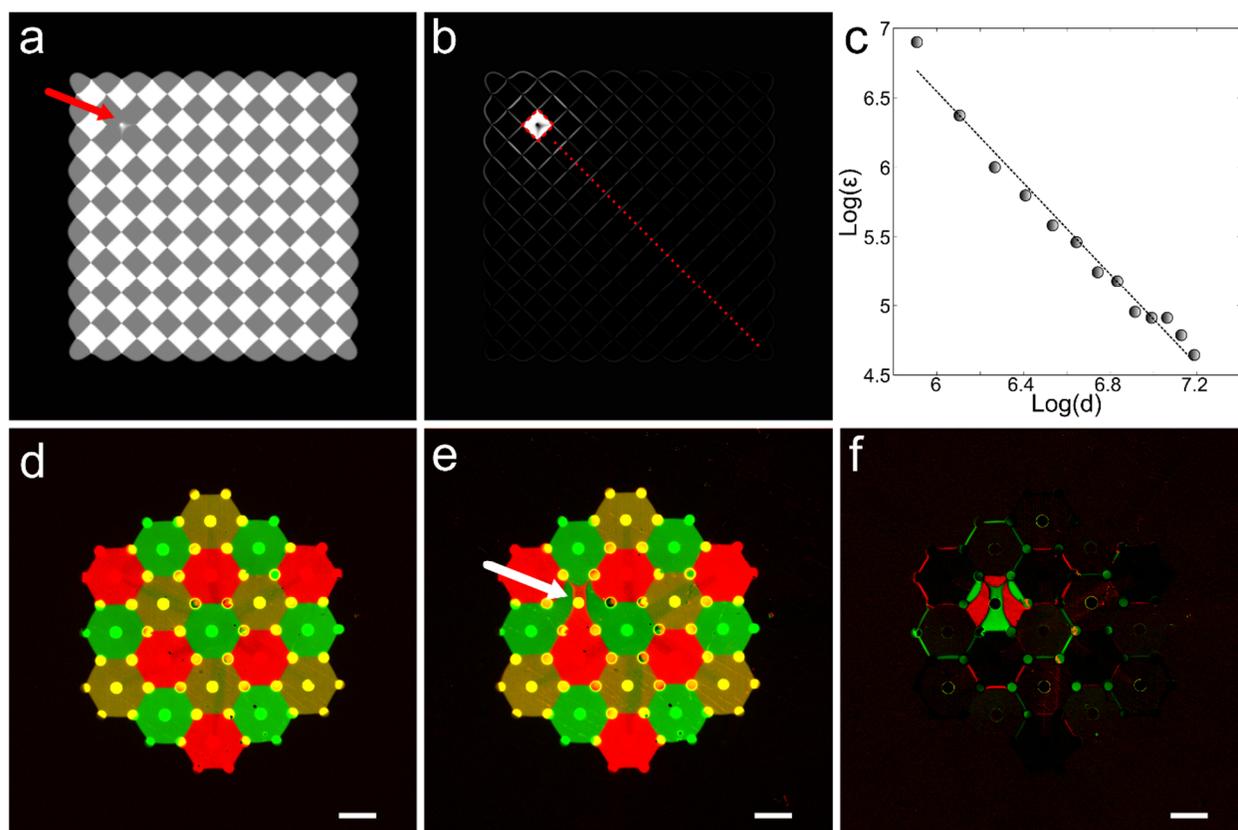

**Fig. 3.** Effect off clogged apertures. (a) Simulation of a 19x19 square-pixel PCD with a clogged aperture. The clogged aperture is pointed by the red arrow. (b) Error resulting from the clogged injection. c) Graph of the error generated at each pixel border by a clogged injection in function of the position. Error were measured on the red line from (b). The error ε follows a decreasing power function of equation $\varepsilon = (1.3 \times 10^7)d^{-1.64}$, where "d" is the distance in pixel. The error (ε) is the integral of errors peak intensity over position. (d) Fluorescent micrograph of a hexagon-pixel PCD. (e) The hexagon-pixel PCD from (d) with an injection turned off. (f) Errors resulting from the injection turned off. Scale bars from (d-f) equal 1 mm.

Next, we investigated the stability of the resulting chemical images projected in the presence of pressure or flow rate mismatches in the display. A classical problem in microfluidics is that flow rate ($Q$) in a split junction is greatly affected by a slight error on the channel hydraulic radius $r$ ($Q \propto r^4$)[20]. To examine this issue, we used a Monte-Carlo method[21]. Approximately 24 000 2D convection-diffusion finite-element transport simulations were performed, where stochastic errors were randomly added to the flow rate of each aperture. By varying the bound of the stochastic errors (from [-5%, 5%] to [-75%, 75%)], the effect on PCDs' stability was observed. The results show that flow rate or pressure errors of up to ±20% have virtually no effect on the PCDs (see SI section 3). Depending on the acceptable pixel deformation threshold for a given application, PCD with flow rate matching errors of up to 40% could be used. Moreover, the PCDs' size (in number of pixels) has no significant effects on its stability or error propagation, as expected from a device based on periodicity (Supplementary Fig. 9). To conclude, while in most systems a higher number of subcomponents lead to more instability, we emphasize here that, in the case of the PCD, the use of periodic streaming flow patterns interacting with each other leads to a stable, self-correcting behavior.

**Reconfigurability**

In turn, this pattern stability indicates that smooth transitions between multiple flow patterns exist and should be readily achievable, suggesting the possibility to stream sequences of chemicals within any given pixel. In the LCD analogy, this is equivalent to enabling a "RGB" like transition in pixels to stream chemical movies onto a processed surface. To demonstrate PCD reconfigurability with a minimum number of pumps and valves, we developed an approach strongly inspired by hardwired relay logic[22]. The apertures injecting the same chemical sequence were grouped into manifolds. We positioned programmable valves upstream of manifolds, so that every chemical image required during an experiment could be formed using a combination of logic operations on the valves (injecting either green fluorophore or red fluorophore) (Fig. 4). An array of programmable syringe pumps, synchronized with the valves, controlled the injected fluids, allowing the generation of sequences of chemical images. We used nine valves in this system, allowing the generation of 512 chemical images, including the letters "P", "O", "L" and "Y" (Fig. 4, Supplementary Video 1). The transition time between different chemical images was in the order of a few tens of seconds. Moreover, the transition time being mainly due to Taylor-Aris dispersion in the tubing, it could be reduced by an order of magnitude or more using smaller tubing or capillaries (SI section 4).

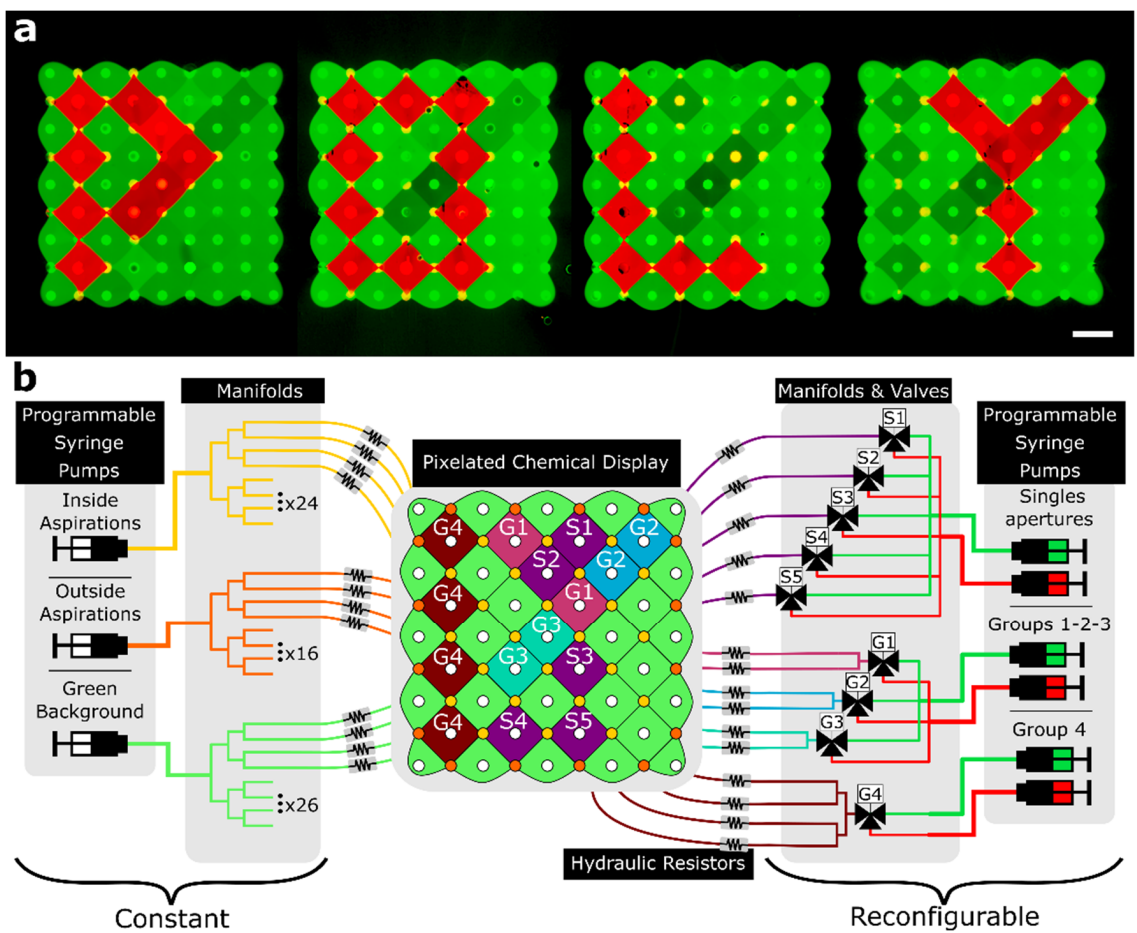

**Fig. 4.** PCD reconfiguration. a) Fluorescence micrographs of PCD reconfigured to sequentially write "P", "O", "L", and "Y". Scale bar represents 1 mm. b) Schematic of the pumps, valves and manifolds used to perform the PCD reconfiguration.

**Roll-to-Roll applications**

Large reconfigurable, stable PCDs open up new possibilities for industrial and fabrication applications. Operating in local confinement in a globally unconfined flow removes the limitation of working in an immersed environment like with previous open-space microfluidics devices. To showcase applications on dry surfaces, we processed thin plastic films in a roll-to-roll setup using the PCD. Two distinct processes were implemented. First, as an example of surface biopatterning, we used the PCD to coat a polystyrene (PS) film simultaneously with two distinct fluorescently labeled forms of bovine serum albumin (BSA-Alexa Fluor 488 and BSA-Alexa Fluor 555). We used an 81-aperture square pixel PCD with fluid pattern similar to Fig. 2d in a custom roll-to-roll setup to pattern the film in a stop and go manner with a cycle of 3 minutes (Fig. 5a and Supplementary Fig. 5). Results show well-defined, sharp patterns without overlap between the two proteins (Fig. 5b). Consecutive patterns could be printed as close as 2.3 mm with the current 3D printed device. Remarkably, the roll-to-roll system integrated with a PCD was found to be insensitive on the gap and small angles between the PCD and the film. This resilience is due to the force equilibrium between the tension in the plastic film and the hydrodynamic pressure generated by the injections of fluids in the narrow gap[23]. We implemented a second roll-to-roll process to print pure silver patterns on a different material: polyethylene terephthalate (PET). For the latter, the objective was to use the thin diffusion layer between reagents in neighboring pixels to generate a controlled interfacial reaction (Fig. 5c). To achieve this result, we used Tollen's silver mirror reaction, a method that already demonstrated its potential for the deposition of silver thin films of controlled thickness for electric circuitry[24,25]. Tollen's Reagent, when mixed with an aldehyde (in this case dextrose), yields solid silver. This reaction result in thin lines of silver being patterned on the PET sheet between pixels injecting Tollen's reagent and aldehyde. Water was injected around pixels where no silver line had to be patterned. To increase the silver deposition speed, a sensitization bath of tin chloride was added prior to the silver deposition[25]. As for protein biopatterning, the process was performed in a stop-and-go manner. The PET film was sensitized in a bath, rolled to the PCD, and then patterned for four minutes and rinsed for a minute while the next surface to be patterned was being sensitized. An example of complex thin film silver patterns (an E-shaped pattern) on PET is demonstrated as an example of controlled interfacial reaction in a PCD (Fig. 5d).

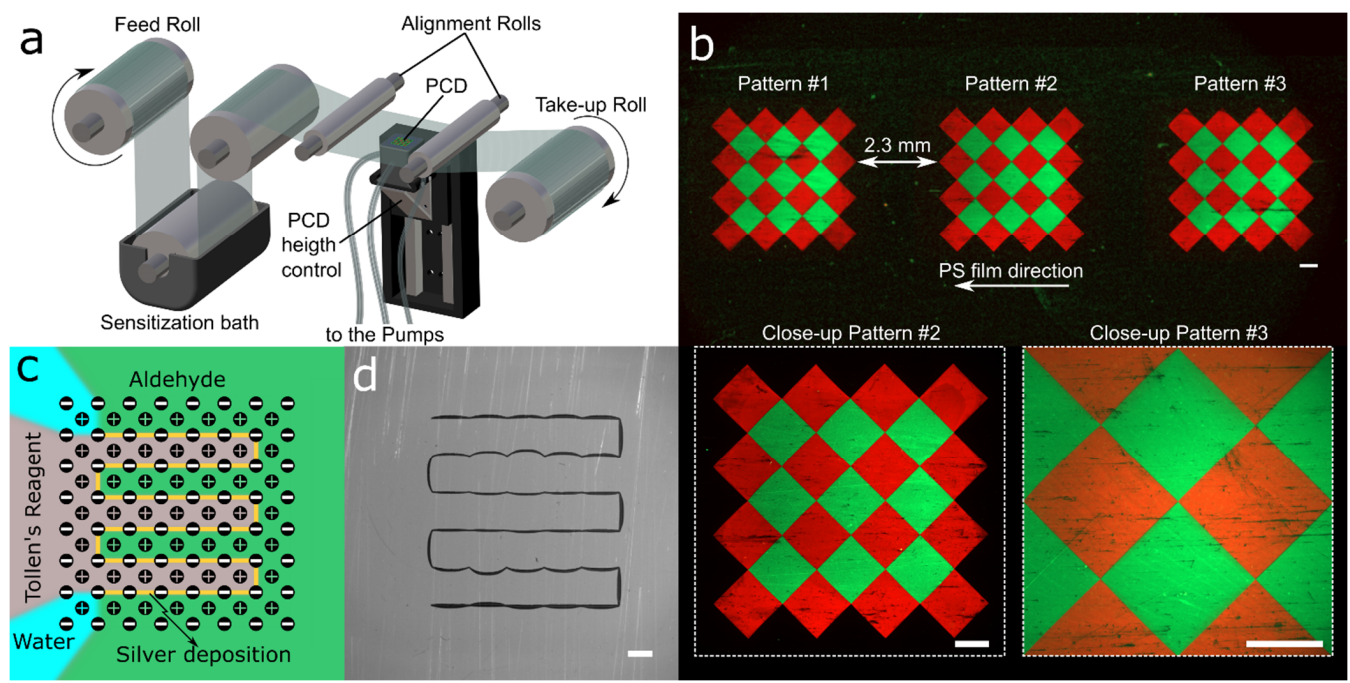

**Fig. 5.** Roll-to-roll processes using a PCD. a) Schematic of the roll-to-roll setup used for the biopatterning of proteins and silver thin film deposition. For the bio-patterning, the sensitization bath was removed, and the film went directly from the feed roll to the alignment rolls b) Fluorescent micrograph of three consecutive patterns of BSA-Alexa 555 and BSA-Alexa 488 proteins on a polystyrene film in a roll-to-roll process. c) Fluids schematic of the deposition process of a E-shape silver pattern on a PET film. d) Micrograph of an E-shape silver pattern on a PET film roll. All scale bars represent 500 μm.

**Conclusion and Outlook**

Pixelated chemical displays offer the first open-space microfluidic solution to process surfaces not immersed in fluids, opening up applications in roll-to-roll printing and thin film technologies. By design, PCDs allow fast reconfiguration, are impervious to evaporation, and are compatible with viscous samples. Moreover, they have no moving parts, which simplifies interfacing with any equipment or experiments requiring live imaging. Emerging 3D printing materials, such as chemically resistant polymers[26] or even glass[27] should allow their compatibility with a wide range of solvents or acids. As presented in this article, PCDs were used for the maskless biopatterning of surfaces and for the deposition of precise silver pattern on thin plastic film. However, they might also find uses in bioassays by allowing the dynamic and multiplexed microperfusion of cell cultures or tissues.

In this article, we produced and operated a PCD of up to 144 pixels (12 x 12) and a reconfigurable system encompassing nine valves, thus allowing the generation of 512 different chemical images. From there, the scaling up of the device to an arbitrary number of pixels is straightforward, since all pixels operate independently from each other and displayed robustness to experimental imperfections. The number of inlets and outlets in a PCD scale linearly with the number of pixels. At least one valve per pixel would also be required to enable independent switching in every pixel as in LCD technology. This is out of consideration using macroscopic switch valves such as those used in this paper. However, recent advances in high resolution 3D printing seem to offer a way forward. Using custom resin and printers, Gong & al. recently showed the possibility to print compact arrays of up to 6 pressure valves per mm$^2$, high-density chip interconnects and channels with a diameter of 20 μm[28] (tenfold smaller than in this article). Using a 3D printer with such resolution could easily reduce the minimum size of microfluidic pixels to around 100 μm and increase pixel density by a factor of 100. Miniature PCDs with integrated valves and manifold systems would greatly cut channel dead volumes and simultaneously, the switching speeds of reagents within pixels. Fueled by the fast developments in 3D printing technologies, PCDs with pixel numbers in the thousands similar in size and resolution to early LCD screens[29] may not be too far away.

## Materials and Methods
**PCDs fabrication and operation**
PCDs were designed using Catia V5 (Dassault Systèmes). They were printed using a stereolithography 3D printer with a x-y resolution of 27 μm and a wavelength of 385 nm (Pico 2 HD, Asiga). The resin used was Pro3dure GR-1 black (Pro3dure GmbH), printed in layers of 50 μm, with an exposure time of 0.468 s. After the printing, PCDs were cleaned in an isopropanol bath in a sonicator. They were then post-cured 10 min in UV light (Flash UV Curing Chamber, Asiga). 1/32" tubes (Masterflex Tygon, Cole-Parmer) were plugged and glued using a UV-sensitive resin (DecorRom). Different experimental setups were used for the figure of this article, in an upright and downright configuration. In some case, PCDs were simply set down on the glass slide, in others, spring or screws were used to hold the PCD in place. The sample distance to the PCD was controlled by the spacer pillars. Flow rates were controlled using nemesys syringe pumps (Cetoni GmbH). For the reconfigurable display, Cetoni 3-way valves were used. Pumps and valves were controlled using a custom LabView code (National Instrument).

**Experimental measurements**
Fluorescent micrographs were taken using an inverted fluorescence microscope (Axio Observer.Z1, Zeiss) with a Lavision sCMOS camera (Lavision), and an upright microscope (Axio Zoom.v16, Zeiss) with a sCMOS camera (Pco.Edge, PCO). Fluorescein salt dissolved in ultrapure water was used as the green fluorophore and propidium iodide (Sigma Aldrich) and DNA sodium salt (DNA sodium salt from salmon testes, Sigma Aldrich) dissolved in ultrapure water was used as the red fluorophore. A background image, which was taken by tuning off the PCD's injections, was subtracted from the measured images to remove the background noise. This process was made using MATLAB. The reconfiguration video was made using the same protocol frame-by-frame before being compressed using MATLAB.

**Simulations**
2D finite element simulation of the PCDs were made using the MATLAB-COMSOL livelink (COMSOL Multiphysics v.5.1, COMSOL AB). Simulations were made by coupling Laplace's equation (velocity potential) with diffusion. Dimensions and flow rates used where in the same order of magnitude as what was used for the experiments. Fluorescein was considered as the diffusive species and the diffusion coefficients used was $0.5 \times 10^{-9}$ m$^2$/s. Pixels with concentration of 1 and 0.5 were simulated to allow the easy segmentation of the pixels. To measure the effect of a clogged injection on a PCD, a 19 × 19 aperture square-pixel PCD with an injection turned off was simulated and compared with a perfectly working PCD. The error measured (ε) was the integral of errors peak intensity over position at each pixel transition.

## Stability Analysis

A MATLAB code was used to automatize simulations with randomized errors on the flow rate on every aperture (each injection having flow rate randomized in the range [Q - (limit(%) × Q), Q + (limit(%) × Q)]). After the simulation, the pixels were segmented using a level threshold. Morphological mathematical image opening (erosion followed by dilatation) was used to improve the segmenting of the microfluidic pixels. For each microfluidic pixel, the area was measured, and the usability threshold criteria was tested. The standard deviation of the pixels' areas and usability threshold were calculated for every simulation and were then averaged over a thousand simulations for each condition. The standard error was calculated over the 1000 simulations. 95% bilateral confidence intervals on the standard deviation of the area and the ratio of pixels respecting the usability threshold were calculated using ±1.96 × (Standard Error). The analysis didn't consider the border pixels of the PCD since they don't have the same geometry as the other pixels. Also, in a small number of cases, for simulation with high flow rate errors ($\geq$ 50%), our algorithm couldn't properly segment clusters of two to three microfluidic pixels. In this case, the area considered for each pixel was the combined area divided by the number of pixels. In the rare case where more pixels were wrongly segmented as one, their value was simply rejected. For each simulation, the segmented microfluidic pixel image was compared to a segmented image of PCD without error on the flow rate and an error map was made. By calculating the proportion of error maps containing an error for each point, the probability of error heatmaps were generated. More details in Supplementary Section 3.

## Roll-to-roll experiment

A custom roll-to-roll setup was built for 8 cm width film. The setup comprised a motorized stage (X-LRM, Zaber) to control the gap in between PCD and the plastic film. The stage was controlled using a custom LabView code. A 190 µm thick polystyrene (PS) film (Goodfellow) was patterned using 10 µg/ml BSA-Alexa Fluor 555 and BSA-Alexa Fluor 488 (Thermo Fisher Scientific) in a solution of 25 mM acetate buffer (Thermo Fischer) at pH 5, chosen to equal the proteins' isoelectric point to improve adsorption on the PS. The film surface was patterned for two minutes at a flow rate of 800 nL/s per aperture. The surface was then rinsed for one minute by stopping the BSA injection. Acetate buffer was injected by the outside injection while rolling the roll to ensure that a film of liquid was always present between the PCD and the film, and thus that no air would be aspirated.

The silver circuit patterning was performed on a 250 µm polyethylene terephthalate (PET) film (McMaster-Carr). The film was sensitized in a bath of 50 mM tin chloride (Sigma Aldrich) for 5 minutes. The tin chloride solution was made by mixing the tin chloride to ultrapure water, and then drops of 1 M HCL (Sigma Aldrich) were added until the dissolution of the tin salt. The patterning was performed by injecting 50 mM Tollen's reagent by a set of apertures, and 50 mM dextrose (Sigma Aldrich) by another set of apertures. Tollen's reagent was made by mixing a 100 mM silver nitrate (Sigma Aldrich) solution in deionized water. 100 mM KOH solution (Sigma Aldrich) was added at a 1:1 ratio. 30% ammonia (Sigma Aldrich) was then added dropwise until the precipitate dissolve[25]. The PET film was patterned for 4 minutes with a flow rate of 800 nL/s per aperture. The film was then rinsed for a minute by stopping the dextrose and Tollen's reagent injection while keeping the deionized water injection and central aspirations running.

## Data Availability

The data and codes associated with the stability analysis are available from the corresponding author on request. All other data are contained in this article and its supplementary information files.

## Acknowledgments

P.A.G. and E.B. acknowledge a graduate scholarship from the Fond de Recherche du Québec Nature et Technologies (FRQNT) and from the Natural Sciences and Engineering Research Council (NSERC). M.T. acknowledges an undergraduate internship scholarship from the NSERC. T.G. acknowledges support from NSERC (RGPIN-06409) and FRQNT (#205993, #197507) grants. We acknowledge CMC Microsystems for the shared access to design and simulation tools. We acknowledge David Juncker, Klavs F. Jensen, Molly M. Stevens, Stéphane Kena-Cohen and Daniel Therriault for useful discussions on the manuscript.

**Author contributions**

P.A.G. and T.G. designed the research. P.A.G. designed and fabricated the PCD and the experimental setups. P.A.G. performed the fluorescence and roll-to-roll experiments. E.B. performed the analytical modeling. M.T. carried out the simulations. M.T. and P.A.G. realized the stability analysis. P.A.G. and T.G. wrote the manuscript. All authors reviewed the manuscript.

**Competing interests**

The authors declare no competing financial interests.